\documentclass[aps,floatfix,prd,twocolumn,nofootinbib]{revtex4}
\usepackage{graphicx}
\usepackage{amsmath}
\usepackage{amssymb}
\usepackage{subfigure}

\newcommand{\MeV}{\;\text{MeV}}

\newcommand{\Gauss}{\;\text{Gauss}}
\newcommand{\B}{{B}}
\newcommand{\jem}{{j_{em}}}
\newcommand{\I}{$I\bar{I}$\;}

\begin{document}

\title{Correction to the Chiral Magnetic Effect from axial-vector interaction}

\author{Zhao~Zhang}\email{zhaozhang@pku.org.cn}
\affiliation{ School of Mathematics and Physics, North China
Electric Power University, Beijing 102206, China}

\begin{abstract}
The recent lattice calculation at finite axial chemical potential suggests that the induced current density of the chiral
magnetic effect (CME) is somehow suppressed comparing with the standard analytical formula. We show in a NJL-type model of QCD
that such a suppression is a natural result when considering the influence of the attractive axial-vector interaction.
We point out that the lattice result doesn't need to be quantitatively consistent with the analytical formula due to the chirality
density-density correlation. We also investigate the nonperturbative effect of instanton molecules on the CME. Since an
unconventional repulsive axial-vector interaction is induced, the CME will be enhanced significantly by the instanton-anti-instanton pairings.
Such a prediction needs to be tested  by more improved lattice simulations. We further demonstrate that the
axial-vector interaction plays an important role on the $T-\mu_A$ phase diagram.
\end{abstract}
\maketitle

\section{Introduction}
\label{sec:Introduction}
The topological charge of gauge field theory is defined as
\begin{equation}
 Q_T = \frac{g^2}{32\pi^2}\int d^4 x \;\text{Tr}[F\tilde{F}],
\end{equation}
where $F$ and $\tilde{F}$ refer to the field strength tensor and its dual, respectively. It is well-known that Quantum Chromodynamics(QCD) contains nontrivial gauge configurations carrying the topological charge with an integer number\cite{ref:instanton}. These configurations interpolate between the topologically different vacua of QCD characterized by different Chern-Simons numbers. For a long time the experimental evidence for the existence of the  topological gluon configurations only comes indirectly from the meson spectrum\cite{ref:meson}.

Recently, the study of the chiral magnetic effect(CME)\cite{Kharzeev:2007jp, Fukushima:2008xe} has received much attentions since the topological configurations of QCD may be observed directly in the noncentral heavy-ion collisions. The essence of the CME is the chiral imbalance induced by the nonzero topological charge through the axial anomaly of QCD
\begin{equation}
 (N_R - N_L)_{t=+\infty} - (N_R - N_L)_{t=-\infty} = -2 Q_T \;,
\label{eq:int1aaa}
\end{equation}
where $N_R(N_L)$ denotes the quark number of the right-handed(left-handed) chirality in the chiral limit. Eq.~(\ref{eq:int1aaa}) indicates the chiral asymmetry is generated by the nonzero topological excitations. In a strong magnetic field which may be produced in the noncentral heavy-ion collisions, a net electric current will be induced along the direction of the magnetic field due to the chirality imbalance.

In order to study the effect of the chiral asymmetry, usually the axial chemical potential $\mu_A$ is introduced in a grand canonical ensemble\cite{Fukushima:2008xe}. The physical meaning of $\mu_A$ is the difference of
the right- and left-handed quark chemical potentials. This quantity can be identified as the time derivative of the $\theta$ angle of QCD, namely $\mu_A=\partial_0{\theta}/{2N_f}$\cite{Fukushima:2008xe}. In the presence of an external magnetic field $\B$, the analytical expression of the induced electric current density for the CME at finite $\mu_A$ has been given in \cite{Fukushima:2008xe} which takes the form
\begin{equation}
\jem=CN_cN_f{e^2{\mu}_A\B},
 \label{eq:jz-zeroga}
\end{equation}
where $C=1/{2\pi^2}$. Note that all quarks carrying the same electric charge is assumed in \eqref{eq:jz-zeroga}. Obviously, this equation explicitly breaks the $\mathcal{P}$- and $\mathcal{CP}$-symmetry.

Equation.(\ref{eq:jz-zeroga}) indicates that the hot quark-gluon plasma(hQGP) is an ideal place to study the physics of $\mathcal{P}$- and $\mathcal{CP}$-odd excitations of QCD. First, the production probability of the configurations with nonzero winding numbers becomes higher since jumping over the potential barrier between distinct classical vacua is probable at high temperatures. The configurations responsible for such thermal transitions are called sphalerons\cite{MMS,BMR} in QCD, which might occur at a copious rate at high temperature compared to the low-temperature tunneling by the instanton. Second, a strong magnetic field can be produced in non-central heavy-ion collisions\cite{Kharzeev:2007jp,Skokov:2009qp}. The study based on the UrQMD model\cite{Skokov:2009qp} suggests that the $e\B$ produced at RHIC and LHC can be as large as $e\B\approx{2m_\pi^2}$ and $e\B\approx{15m_\pi^2}$, respectively\footnote{The magnetic field strength can be translated into the CGS system with the identity $m_\pi^2\approx 10^{18} \Gauss$, where $m_\pi=140\MeV$.}. So in a noncentral heavy-ion collision, the induced electric current may lead to an excess of the positive electric charge on one side of the reaction plane and the negative electric charge on the other. Such a charge separation effect may be observed experimentally. Recently, a conclusive observation of charge azimuthal correlation has been presented by the Star collaboration \cite{STAR} which may result from the CME with local $\mathcal{P}$- and $\mathcal{CP}$-violation.

Theoretically, the investigation of the CME has been performed recently in the Lattice QCD simulation by introducing a finite $\mu_A$\cite{Yamamoto:2011gk}. The advantage of the lattice simulation at finite axial chemical potential is that there is no sign problem comparing with the case at finite baryon chemical potential. In this study, the CME described by Eq.\eqref{eq:jz-zeroga} has been confirmed qualitatively. But at the quantitative level, the obtained electric current density is significantly suppressed \cite{Yamamoto:2011gk} compared to the analytical formula (\ref{eq:jz-zeroga}). The further study at the quenched level of QCD suggests that the lattice data are sensitive to the lattice spacing\cite{Yamamoto:2011ks} and the obtained data increase significantly compared to the ones in\cite{Yamamoto:2011gk}. However, the deviation of the new lattice data from the analytical formula is still sizable even some systematic errors have been taking into account\cite{Yamamoto:2011ks}.

The suppression of the lattice data may stem from some errors of the lattice simulation and/or the non-perturbative effects of QCD. Accordingly, besides taking more improved lattice calculations, physically understanding the current lattice result through other methods of QCD is very important and necessary. On the other hand, due to the limitation of the lattice calculation, investigating the non-perturbative corrections and making new predictions on the CME are also very required by using the effective theories or models of QCD.

The purpose of this paper has two aspects: First, we will try to disclose the possible mechanism for the deviation of the present lattice data from the analytical result by taking into account some QCD corrections; Second, we will explore the non-perturbative influence of the instanton molecules on the CME and make some predictions. Note that both aspects are closely related to the effective four-quark interaction in the axial-vector channel, where the chirality density-density correlation plays an important role.

The paper is organized as follows: In Section \ref{sec:Effective}, we give the effective quark interactions of QCD for $T>T_c$. In section \ref{sec:Correction}, we present the correction to the CME from the axial-vector interaction. Section \ref{sec:Numerical} is devoted to the numerical results and related discussions. The last section is the summary and conclusion.

\section{Effective quark interactions for $T>T_c$}
\label{sec:Effective}

The experimental data of RHIC suggests that strongly coupled quark-gluon plasma(sQGP) is formed for $T>T_c$, which exhibits the behavior of a perfect fluid. Now it is widely believed that the non-perturbative QCD still plays very important roles on the hQGP up to $2-3T_c$. Since the chiral magnetic effect may be observed directly in heavy-ion collisions, one naturally expects that the role of the non-perturbative aspects of QCD on the CME should be quite remarkable. One of the possible non-perturbative effects may arise from the instanton-anti-instanton (\I) molecules at finite temperature near $T_c$.

As a major component of the QCD vacuum, the instantons can produce the quark condensate, the low-lying hadron state and other non-perturbative features of QCD, according to the interacting instanton liquid model (IILM) \cite{Schafer:1996wv}. At zero and low temperatures, the instanton ensemble is in a random liquid state which is responsible for the spontaneous chiral symmetry breaking and the axial anomaly. Naively one would expect that the instantons might give a little contribution to the CME due to their significant suppression at high temperature.

However, it has been suggested that the hot chiral transition is not driven by the vanishing of the instantons and anti-instantons but by their rearrangement around $T_c$ \cite{ilgenfritz1,Ilgenfritz:1994nt,Schafer:1994nv}. Namely, when approaching to $T_c$ from below, the random instantons and anti-instantons are paired up into the ordered \I molecules. This picture implies that the instantons still keeps sizable density near $T_c$ with the molecular forms. Moreover, a recent investigation indicates that the \I molecules can even survive up to $\sim5T_c$\cite{Wantz:2009mi}.

The rearrangement picture indicates that the non-perturbative effects induced by the instanton configurations will persist into the high temperature region. In the chiral symmetric phase, the interacting \I molecules can be regarded as one of the possible mechanisms responsible for the sQGP formed in $T_c<T<2\sim3T_c$\cite{Brown:2003km}. So even the chiral restoration is one of the necessary conditions for the appearance of the electric current $\jem$, the instantons may still exert important influence on the chiral magnetic effect via the molecular forms.

It is well known that the random instanton configurations can lead to the famous 't Hooft interactions among quarks with different flavors
\begin{equation}
\mathcal{L}_{'t Hooft} \sim \prod_{f}(\bar{\psi}\phi_0)(\bar{\phi}_0\psi),
\label{eq:thooft}
\end{equation}
which explicitly breaks the $U_A(1)$ symmetry due to the axial anomaly. Similarly, the ordered \I  molecules also result in effective quark interactions, but only with the four-fermion coupling forms. For two-flavor case, the Fierz-invariant quark interactions induced by \I molecules has been derived in \cite{Schafer:1994nv} with the form
\begin{align}
 \mathcal{L}_{mol sym}&=G\Big\{\frac{2}{N^2_{c}}\sum_{a=0}^{3}[(\bar{\psi}\tau^{a}\psi)^2-
 (\bar{\psi}\tau^{a}\gamma_5\psi)^2]\nonumber\\
 &+\frac{2}{N^2_{c}}(\bar{\psi}\gamma_{\mu}\gamma_5\psi)^2-\frac{1}{2N^2_{c}}\sum_{a=0}^{3}[(\bar{\psi}\tau^{a}\gamma_{\mu}\psi)^2\nonumber\\
 &+(\bar{\psi}\tau^{a}\gamma_{\mu}\gamma_5\psi)^2]
 \Big\}+\mathcal{L}_8,
\label{eq: molecule}
\end{align}
where $\tau^{0}$ and $\vec{\mathbf{\tau}}$ are unit and Pauli matrices in flavor space respectively and $G$ is the coupling constant. The $\mathcal{L}_8$ in \eqref{eq: molecule} refers the quark interactions in the color octet channel. In the \I molecular picture, it is expected that the dominant quark interactions for $T>T_c$ are the molecular ones rather than the 't Hooft interactions \cite{Ilgenfritz:1994nt,Schafer:1994nv,Wantz:2009mi,Brown:2003km}\footnote{Note that the 't Hooft type interactions might also survive in the $T>T_c$ region according to \cite{Wantz:2009mi}. Such interactions should have little influence on the CME since the electric current only appears in the chirally symmetric phase.}. Different from (\ref{eq:thooft}), the lagrangian \eqref{eq: molecule} respects not only the $SU_V(2)\otimes{SU_A(2)}\otimes{U_B(1)}$ symmetry but also the ${U_A(1)}$ symmetry. This is because the \I pair gives vanishing topological charges\cite{Schafer:1994nv}.

Note that the most general Fierz-invariant form for the four-quark interactions under the invariance of $SU_V(2)\otimes{SU_A(2)}\otimes{U_B(1)}\otimes{U_A(1)}$ symmetry including both the scalar(psudo-scalar) and vector(axial-vector) currents is
\begin{align}
 \mathcal{L}^{(4)}_{general}&=\frac{1}{2}G_1\sum_{a=0}^{3}[(\bar{\psi}\tau^{a}\psi)^2+(\bar{\psi}\tau^{a}i\gamma_5\psi)^2]\nonumber\\ &-\frac{1}{2}G_2\sum_{a=0}^{3}[(\bar{\psi}\tau^{a}\gamma_{\mu}\psi)^2+(\bar{\psi}\tau^{a}\gamma_{\mu}\gamma_5\psi)^2]\nonumber\\
 &-\frac{1}{2}G_3[(\bar{\psi}\tau^{0}\gamma_{\mu}\psi)^2+(\bar{\psi}\tau^{0}\gamma_{\mu}\gamma_5\psi)^2]\nonumber\\
 &-\frac{1}{2}G_4[(\bar{\psi}\tau^{0}\gamma_{\mu}\psi)^2-(\bar{\psi}\tau^{0}\gamma_{\mu}\gamma_5\psi)^2]+\mathcal{L}^{(4)}_{8},
 \label{eq:full4f}
\end{align}
where $G_i$ are four independent coupling constants \cite{Klimt:1989pm}.  Comparing \eqref{eq: molecule} with \eqref{eq:full4f}, one can find that the interacting $I{\bar{I}}$ molecule model (IIMM) gives prediction of these coupling constants with
\begin{align}
G_1=\frac{4}{N_c^2}G,\quad G_2=\frac{1}{N_c^2}G,\quad G_4=-G_3=\frac{2}{N_c^2}G,
 \label{eq:coupling}
\end{align}
which are all dependent on the only parameter $G$. We shall show below that nonzero $G_2$ and $G_{3(4)}$ give nontrivial contributions to the induced current $\jem$ for the CME.

We stress that even we are very interested in the possible non-perturbative influence of the instantons on the CME, we do not limit our study  within the only formalism \eqref{eq: molecule}, which is just a special case of \eqref{eq:full4f}. Without loss of generality, we will assume that the Lagrangian density with the four-quark interactions \eqref{eq:full4f} works for $T_c<T<2\sim3T_c$. This Lagrangian can be regarded as an extension of the traditional NJL model of QCD (In the next section, the degree of freedom of the Polyakov-loop is also included).

For the physics related to the CME, we are particularly interested in the effective four-quark interactions in the vector isospin-scalar and axial-vector isospin-scalar channels
\begin{equation}
 \mathcal{L}_{VA} = -G_V (\bar{\psi}\gamma_\mu \psi)^2
  -G_A (\bar{\psi}\gamma_\mu\gamma_5 \psi)^2 \;,
\label{eq:va-interaction}
\end{equation}
which appears in both \eqref{eq: molecule} and \eqref{eq:full4f}.
The significance of $\mathcal{L}_{VA}$ to the CME is attributed to the appearance of two special vacuum expectation values (VEVs) in the chirally  asymmetric system under the influence of an external magnetic field $\B$, namely, $j_z=\langle{\bar{\psi}\gamma^3}\psi\rangle$ and $n_A=\langle{\bar{\psi}\gamma_{0}\gamma_5}\psi\rangle$\cite{Fukushima:2008xe}. The former is the VEV of the vector current $\bar{\psi}\gamma^\mu{\psi}$ along the direction of $\B$ and the latter the chiral charge density conjugate to $\mu_A$. As a consequence, the chiral magnetic effect might be sensitive to the couplings $G_{V}$ and $G_{A}$ in \eqref{eq:va-interaction}.

The prediction of the IIMM gives
\begin{equation}
G_V=\frac{G_2}{2}=\frac{G}{2N_c^2},\ G_A=2G_4+\frac{G_2}{2}=-3\frac{G}{2N_c^2}.
\label{eq:gvgaIIMM}
\end{equation}
Equation \eqref{eq:gvgaIIMM} indicates two unconventional points induced by the \I molecules: (1) The vector interaction is attractive while the negative $G_A$ implies that the axial-vector interaction is repulsive; (2) The magnitude of $G_A$ is three times that of $G_V$ (For the totally polarized case, the coupling $G_A$ in the temporal direction is even 12 times that of $G_V$\cite{Schafer:1994nv}).

This is quite different from the conventional four-quark interactions derived from the Fierz transformation of the colored quark-antiquark current-current interaction
\begin{equation}
\mathcal{L}_{OGEA}=g(\bar{\psi}\gamma_\mu \lambda^a_c\psi)^2,
\end{equation}
which arises from the one gluon exchange approximation (OGEA) of QCD. In the OGEA, both $G_V$ and $G_A$ are positive with
\begin{equation}
G_V=G_A=\frac{G_S}{2},
\label{eq:gvgaGCM}
\end{equation}
where $G_S$ is the coupling constant in the scalar iso-scalar channel.

Note that it is very likely that the effective quark interactions stemming from the OGEA can persist into the non-perturbative region of QCD with relatively strong couplings. A well-known example in the literature is the global color model (GCM) of QCD, which is based on the nonlocal colored quark-antiquark current-current interaction
\begin{equation}
\mathcal{L}_{GCM}(x,y)=g(x-y)\bar{\psi}(x)\gamma_\mu \lambda^a_c\psi(x)\bar{\psi}(y)\gamma^\mu \lambda^a_c\psi(y),
\end{equation}
where $g(x-y)$ is a function of the strong coupling\cite{Cahill:1985mh}.
This model can successfully describe the low-lying hadrons\cite{Tandy:1997qf} and give reasonable QCD condensates in the vacuum and at finite temperature and density\cite{Zhang:GCM}.

Since the axial-vector interaction may be repulsive (as in the IIMM) or attractive (as in the OGEA or GCM), we will treat the coupling $G_A$ in \eqref{eq:va-interaction} as a free parameter. The realistic $G_A$ may include contributions from both the \I molecules and the OGEA, which should depend on the temperature in general. On the other hand, since the vector interaction is attractive in both the IIMM and GCM, we only consider  the positive $G_V$ in the following.

\section{Correction to the CME from Axial-vector Interaction}
\label{sec:Correction}

Working at the mean field level, the Lagrangian \eqref{eq:va-interaction} can be rewritten as
\begin{equation}
\mathcal{L}_{VA}^{MF}=G_{A}n_A^2-2G_An_A\bar{\psi}\gamma_0\gamma_5\psi-G_V j^{z2}+2G_V j^z \bar{\psi}\gamma^3\psi.
\label{eq:meanfield}
\end{equation}
We can read two nontrivial points from Eq.\eqref{eq:meanfield}. First, both squared terms of the condensates $j^z$ and $n_A$ give contributions to the thermodynamical potential. Second, the four-quark interactions in the axial-vector and vector channels lead to a dynamical axial chemical potential $\mu'_A=-2G_An_A$, and an effective gauge field $A^z=-2G_Vj^z$ along the direction of $\B$, respectively. Note that both points are not taken into account in \cite{Fukushima:2008xe} to derive the analytical formula for the CME current.

To investigate the dynamical influence of $\mathcal{L}_{VA}$ on the CME, we add the kinetic lagrangian for free-interacting quarks, namely $\mathcal{L}_{kin} = \bar{\psi} \bigl(i\gamma_\mu \partial_{\mu}+\mu_A \gamma^0\gamma^5 \bigr)\psi$ to the Lagrangian \eqref{eq:full4f}. We will work in the chiral limit throughout this paper. To describe the confinement-deconfinement transition of QCD, we also include the dynamics of the Polyakov-Loop, the action of which in the pure gauge theory denotes as $\mathcal{U}$.  Our model can be looked as the Polyakov-loop-enhanced IIMM but with a varying axial-vector coupling $G_A$. In form, it is very similar to the so called two-flavor PNJL model\cite{PNJL}.

One then gets the mean field thermodynamical potential
\begin{align}
  \Omega & =\mathcal{U}+G_S \sigma^2+G_V j^{z2}-G_A n_A^{2} \nonumber\\
  &-N_c\sum_{f=u,d}\frac{|q_fB|}{2\pi} \sum_{s,k}\alpha_{sk}
  \int_{-\infty}^\infty \frac{d p_z}{2\pi} \,
  f_\Lambda^2 \, \omega_s(p) \notag \nonumber\\
 & -2T\sum_{f=u,d}\frac{|q_fB|}{2\pi} \sum_{s,k} \alpha_{sk}
  \int_{-\infty}^\infty \frac{d p_z}{2\pi} \nonumber\\
 & \times \ln\bigl( 1+3\,\Phi e^{-\beta \omega_s}
  + 3\,\bar\Phi e^{-2\beta \omega_s} + e^{-3\beta \omega_s} \bigr)\;,
\label{eq:omega}
\end{align}
where
\begin{equation}
 \alpha_{sk} = \left\{ \begin{array}{ll}
  \delta_{s,+1} & \text{ for } \quad k=0,\;\; eB>0  \;,\\
  \delta_{s,-1} & \text{ for } \quad k=0,\;\; eB<0  \;,\\
  1            & \text{ for } \quad k\neq0          \;,
 \end{array} \right.\label{eq:index}
\end{equation}
and $f_\Lambda$  is the UV regulator function. $\Phi$ and $\bar\Phi$ in \eqref{eq:omega} correspond to the normalized traced Polyakov loop and its Hermitean conjugate respectively. In \eqref{eq:index}, $s$ refers to the spin and $k$ denotes the Landau level.  The dispersion relations of the quasi-particles derived from the eigenvalues of the Dirac operator take the form
 \begin{equation}
 \omega_s^2 = M^2 + \bigl[ |{\mathbf{p}}| + s\,\tilde{\mu}_A \text{sgn}(p_z) \bigr]^2 \;,
\label{eq:dispersion}
\end{equation}
where
\begin{equation}
|\mathbf{p}|^2 =|\tilde{p}^z|^2 + 2|q_fB|k
\end{equation}
and $q_f$ stands for the electric charge for $f$(u or d) quark. In the above equations, $\tilde{\mu}_A$ and $\tilde{p}^z$ refer to the effective axial chemical potential and the modified momentum along $\mathbf{B}$ respectively, namely
\begin{align}
\label{eq:effua} \tilde{\mu}_A=\mu_A+\mu'_A=\mu_A-2G_An_A,\\
\label{eq:effaz} \tilde{p}^z=p^z+A^z=p^z-2G_Vj^z.
\end{align}
The current $\boldsymbol{j}^z$, the chirality density $\boldsymbol{n}_A$, the quark condensate\footnote{Strictly speaking, there should be two chiral order parameters in the chiral limit since $\langle\bar{\psi}_L\psi_R\rangle\neq\langle\bar{\psi}_R\psi_L\rangle$ for nonzero $\mu_A$.} $\sigma$ and the VEV of the Polyakov-loop $\Phi$ can be determined self-consistently by solving the saddle point equations
\begin{align}
\frac{\partial\Omega}{\partial{n_A}}=0,\quad
\frac{\partial\Omega}{\partial{j^z}}=0,\quad
\frac{\partial\Omega}{\partial{\sigma}}=0,\quad
\frac{\partial\Omega}{\partial{\Phi}}=0.
\label{eq:gapeqs}
\end{align}

Note that the influence of $G_V$ on the CME has been investigated in \cite{Fukushima:2010zza}, where only the vector interaction is introduced phenomenologically. The main conclusion is that $G_V$ may lead to a dielectric correction to the CME current density
\begin{equation}
j = \frac{1}{2{\pi}^2(1+2G_V \mathcal{C}_R)} \cdot N_cN_f e {\mu_A \B} \;,
\label{eq:jrengv}
\end{equation}
where $\mathcal{C}_R = \frac{|eB|}{2\pi^2}$ \cite{Fukushima:2010zza}. When including the axial-vector interaction, the dispersion relation \eqref{eq:dispersion} suggests that Eq.\eqref{eq:jrengv} will be further modified as
\begin{equation}
j = \frac{1}{2{\pi}^2(1+2G_V \mathcal{C}_R)} \cdot N_cN_f e {\tilde{\mu}_A \B} \;.
\label{eq:jrengvga}
\end{equation}

However, a subtlety on the calculation of the polarization tensor or susceptibility in \cite{Fukushima:2010zza} has been pointed out by Kenji Fukushima \cite{Fukushima:2011nu}: According to Eqs.(8-10) in \cite{Fukushima:2011nu}, the quantity $\mathcal{C}_R$ should be zero since the corresponding static susceptibility $\Pi^{33}(q^0=0,\vec{q}\rightarrow 0)$ is always vanishing. This implies that the vector interaction does not give any correction to the CME. Actually, this point is also consistent with the Nielsen-Ninomiya's argument \cite{Nielsen:1983rb} which is only based on the nonrenormalization of triangle anomalies and the energy conservation. As shown in Eq.\eqref{eq:jrengv} (or Eq.\eqref{eq:jrengvga}), the nonzero $\mathcal{C}_R$ will lead to a modification of the coefficient before $\mu_A \B$ (or $\tilde{\mu}_A \B$ in our case). Nevertheless, the argument proposed by Nielsen and Ninomiya \cite{Nielsen:1983rb} does not support the coefficient receiving any correction. This is because such a coefficient is determined by the the nonrenormalization of triangle anomalies, which is  known exactly due to the topological nature of anomalies (more details of this argument see the first part of Sec. III in \cite{Fukushima:2008xe} ).

In light of this, the correct formula for the current density should be
\begin{equation}
{j} = \frac{1}{2{\pi}^2}N_cN_fe\tilde{\mu}_A \B\;
\label{eq:jrenga}
\end{equation}
rather than Eq.\eqref{eq:jrengvga} when considering the axial-vector interaction. We note that, unlike Eqs.\eqref{eq:jrengv} and \eqref{eq:jrengvga}, Eq.\eqref{eq:jrenga} does not contradict Nielsen-Ninomiya's argument. This is because the energy required to remove a particle from the left-handed Fermi-surface and add it to the right-handed Fermi-surface is no longer $2{\mu}_A $ but $2\tilde{\mu}_A$ due to the influence of the axial-vector interaction. In fact, Eq.\eqref{eq:jrenga} will be obtained naturally if we take the same method as part A of Sec. III in \cite{Fukushima:2008xe} but with the replacement of ${\mu}_A $ with $\tilde{\mu}_A$.  

In the following, we will ignore the vector interaction and only consider the effect of the axial-vector interaction on the CME. Using the method
\begin{equation}
 \jem = -\frac{\partial\Omega}{\partial A_3}\biggr|_{A_3=0}\;
\label{eq:numDer}
\end{equation}
and the replacement
\begin{equation}
\frac{\partial}{\partial A^z} \to q_f \frac{d}{d p_z}\;,
\label{eq:der2}
\end{equation}
one can get the integral expression for the calculation of the electric current density in the two-flavor case
\begin{equation}
 \jem = N_c \sum_{f=u,d} q_f\frac{|q_f B|}{2\pi} \sum_{s,k}\alpha_{ks}
  \int_{-\infty}^{\infty}\frac{dp_z}{2\pi}\frac{d}{dp_z}
  \left[ f_\Lambda^2  \omega_s(p) + \cdots \right] \;.
 \label{eq:jemint}
\end{equation}
As the case of zero $G_A$, only the lowest Landau level (LLL) gives a non-vanishing contribution in \eqref{eq:jemint}.
Ignoring the correction from $G_V$, the induced electric current density $\jem$ for the two-flavor case under the influence of $G_A$ becomes
\begin{equation}
\jem = N_c\sum_{f=u,d}\frac{q_f^2\tilde{\mu}_A\B}{2\pi^2}
  = \frac{5\tilde{\mu}_A{e^2}\B}{6\pi^2} \;.
\label{eq:jz2f}
\end{equation}
As it should be,  Eq.\eqref{eq:jz2f} indicates that the induced electric current density is ultraviolet finite.

Unlike the analytical formula \eqref{eq:jz-zeroga}, the chirality density $n_A$ appears directly in Eq.~\eqref{eq:jz2f}. This density can be obtained self-consistently by
\begin{equation}
\frac{\partial\Omega}{\partial \boldsymbol{n}_A}=0 \;\rightarrow\quad
\boldsymbol{n}_A= -\frac{\partial \Omega}{\partial{\mu_A}}
=-\frac{\partial \Omega}{\partial{\tilde{\mu}_A}} \;,
\label{eq:density}
\end{equation}
according to \eqref{eq:effua}. This equation is similar to the determination of the baryon number density at finite $\mu$ when considering the vector interaction in the NJL-type model\cite{Asakawa:1989bq}. We note that even the formula \eqref{eq:jz2f} is obtained from the LLL integration, it doesn't mean that $\jem$ only depends on the LLL for nonzero $G_A$ since the high Landau levels also give contributions to the chirality density $n_A$.

The modified expression of $\jem$ implies that the induced electric current density under the influence of the axial-vector interaction may deviate from the analytical result in the following aspects. First, the current density $\jem$ is no longer directly proportional to the axial chemical potential $\mu_A$ and the magnetic field $\B$. The reason is that in general, the dynamical chemical potential $\mu'_A$ doesn't linearly depend on $\mu_A$ and $\B$ . Second, $\mu'_A$ is $T$-dependent (Since $G_A$ and $n_A$ are both $T$-dependent), so the current density $\jem$ also depends on the temperature. Third, the IIMM predicts that the CME will be enhanced significantly: (1) $\mu'_A$ has the same sign as $\mu_A$ because of the negative $G_A$; (2) $\mu'_A$ may be sizable since $G_A$ is strong as $G_S$. In contrast, the OGEA (or GCM) predicts that the CME will be weakened by the attractive axial-vector interaction compared to the analytical result.

\section{Numerical Results and Discussions}
\label{sec:Numerical}

In this section, we provide numerical results for the influence of the axial-vector interaction on the chirally imbalanced matter in a constant magnetic field. The coupling $G_S$, the UV regulator function $f_\Lambda(p)$ and the Polyakov-loop model $\mathcal{U}$ in (\ref{eq:omega}) are all adopted from \cite{arXiv:1003.0047}, in which the PNJL model was used to study the CME. The model parameters of the NJL part are fitted by reproducing the pion decay constant and the quark condensate in the vacuum. The $T_c$ for vanishing $\mu_A$, $\B$, and $G_A$ is $\sim228 \MeV$ with the influence of the Polyakov-loop dynamics\cite{arXiv:1003.0047}.

\subsection{Effect of axial-vector interaction on the chiral magnetic effect}
We will focus on the variation of the ratio
\begin{equation}
R={\jem}/{\jem(G_A=0)}={\tilde{\mu}_A}/{\mu_A}
\end{equation}
with the impact of the coupling $G_A$. The quantity $R=R(G_A,T,\mu_A,B)$ corresponds to the ratio $C/2\pi^2$, which is the focus of the recent lattice QCD calculations \cite{Yamamoto:2011gk,Yamamoto:2011ks}. The deviation of $R$ from the unity reflects the influence of the axial-vector interaction on the CME compared to the analytical formula.
\begin{figure}[t]
\begin{center}
\includegraphics[width=0.9\columnwidth]{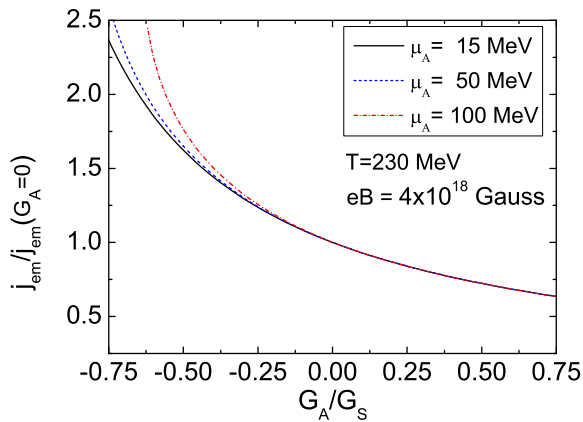} \vspace{1em}\\
\end{center}
\caption{\label{Fig:1}The ratio of $j_{em}$ to $j_{em}(G_A=0)$ as a function of $G_A$ for different $\mu_A$.
The $eB$ and $T$ are fixed as ${4\times10^{18} \Gauss}$ and $230 \MeV$, respectively.}
\end{figure}

In Fig.\ref{Fig:1}, we first show the dependence of $R$ on the coupling $G_A$ for different $\mu_A$ with fixed $eB={4\times10^{18} \Gauss}$ and $T=230 \MeV$. The density of \I molecules might be sizable at this temperature since it is just slightly greater than $T_c$ (The catalytic effect of the magnetic field on the chiral symmetry breaking can be ignored for such an $eB$ \cite{arXiv:1003.0047}). We see that $R$ decreases monotonically with increasing $G_A$. The electric current density is enhanced by $G_A<0$ and weakened by $G_A>0$. Its deviation from the analytical value is quite significant: For the moderate coupling $G_A/G_S=-0.5$ ($0.5$), the electric current density increases (decreases) by $\sim$62 (27) percent at $\mu_A=15 \MeV$; While for the strong coupling $G_A/G_S=-0.75$ ($0.75$), the electric current density increases (decreases) by $\sim$140 (34) percent at the same $\mu_A$. This is in contrast to the minor modification of $\jem$ caused by the vector interaction\cite{Fukushima:2010zza}.
\begin{figure}[t]
\begin{center}
\includegraphics[width=0.9\columnwidth]{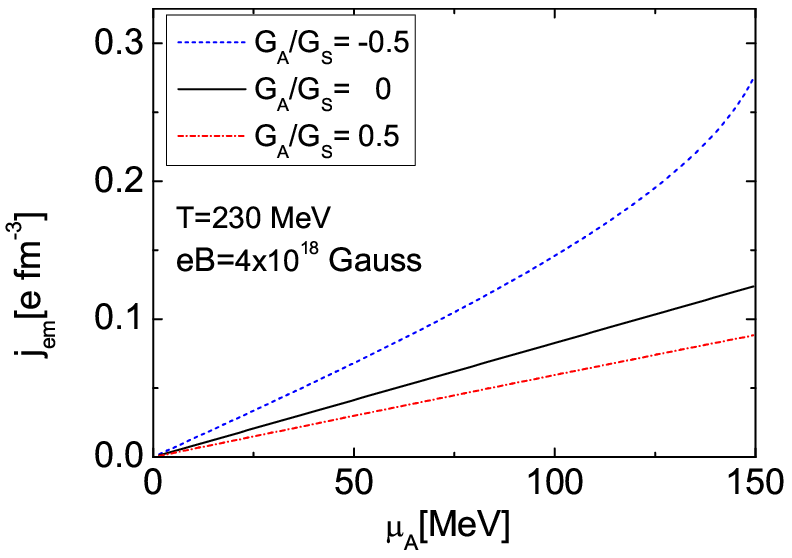} \vspace{1em}\\
\includegraphics[width=0.9\columnwidth]{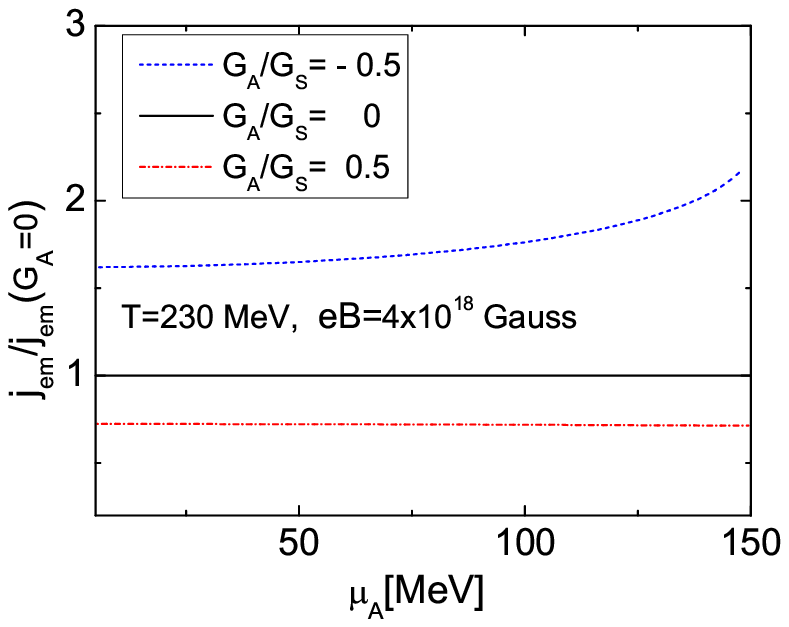} \vspace{1em}
\end{center}
\caption{\label{Fig:2} Upper panel : The current $j_{em}$ as a function of $\mu_A$ for different $G_A$.
Lower panel: The ratio of $j_{em}$ to $j_{em}(G_A=0)$ as a function of $\mu_A$ for different $G_A$.
The $eB$ and $T$ are fixed as ${4\times10^{18} \Gauss}$ and $230 \MeV$, respectively.}
\end{figure}

Figure.\ref{Fig:2} shows the electric current density $\jem$ and the ratio $R$ as functions of $\mu_A$ for different $G_A$, where the same $T$ and $\B$ are used as in Fig.\ref{Fig:1}. For $G_A=-0.5$, the upper panel indicates that the electric current density is nearly proportional to $\mu_A$ in the small $\mu_A$ region, but its deviation from the linear behavior becomes more and more significantly with increasing $\mu_A$. Different from this, the electric current density almost linearly depends on $\mu_A$ up to the moderate $\mu_A$ region for $G_A=0.5$. These features are demonstrated more clearly in the lower panel of Fig.\ref{Fig:2}:  The ratio $R$ monotonically increases with $\mu_A$ for $G_A=-0.5$ while almost remains as a constant for $G_A=0.5$. The same information can be read from Fig.\ref{Fig:1}. This is because the chirality density rises more rapidly with growing $\mu_A$ for the repulsive axial-vector interaction.

\begin{figure}[t]
\begin{center}
\includegraphics[width=0.9\columnwidth]{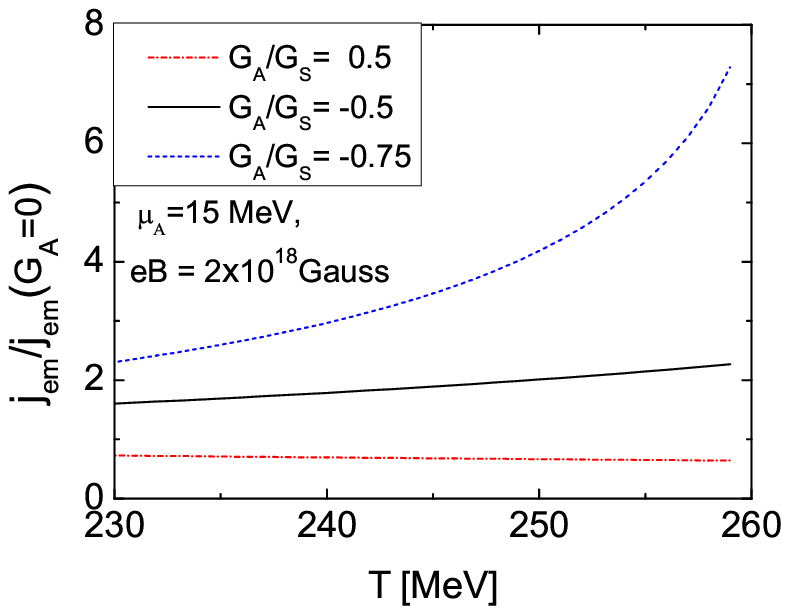} \vspace{1em}\\
\end{center}
\caption{\label{Fig:3} The ratio of $j_{em}$ to $j_{em}(G_A=0)$ as a function of $T$  for
different $G_A$. The $\mu_A$ and $eB$ are fixed as $15 \MeV$ and ${2\times10^{18} \Gauss}$,
respectively. }
\end{figure}

The dependence of $R$ on the temperature with $\mu_A=15\MeV$ and $eB=2\times10^{18} \Gauss$ is displayed in Fig.\ref{Fig:3}. For $G_A<0$, the ratio $R$ monotonically increases with increasing $T$ and its deviation from one becomes huge for the strong coupling. For $G_A>0$, the ratio $R$ decreases relatively slowly with growing $T$ for both the moderate and strong couplings. Though the change of $R$ by $T$ is not so significant as $G_A<0$, the ratio $R$ still decreases by $\sim$10 (15) percent from $\sim{T_c}$ to $\sim{1.1T_c}$ for $G_A/G_S=0.5$ ($1$). Hence, the temperature has a noticeable effect on the induced current. The reason is that the density $n_A$  is a monotonically increasing function of $T$, as reported in \cite{arXiv:1003.0047}.

\begin{figure}[t]
\begin{center}
\includegraphics[width=0.9\columnwidth]{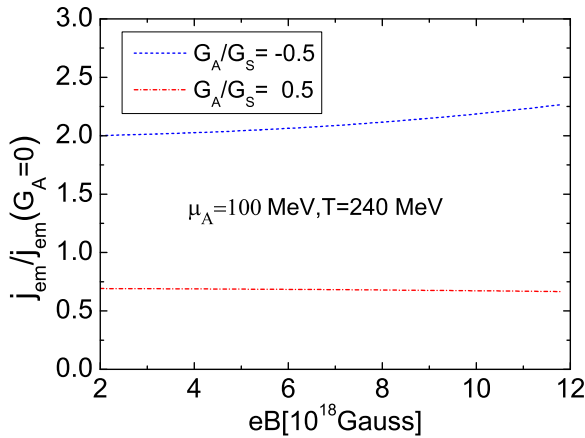} \vspace{1em}\\
\end{center}
\caption{\label{Fig:4}The ratio of $j_{em}$ to $j_{em}(G_A=0)$ as a function of $eB$  for
different $G_A$. The $\mu_A$ and $T$ are fixed as $100 \MeV$ and $240 \MeV$,
respectively.}
\end{figure}

Figure.\ref{Fig:4} indicates that the ratio $R$ is not so sensitive to the external magnetic field but still slowly rises(drops) with $eB$ for $G_A<0$($G_A>0$). This is because that the chirality density is enhanced by the magnetic field through the increase of the thermodynamical potential\cite{Fukushima:2008xe}. Note that such a tendency will slow down or cease for large enough $B$ since its catalytic effect on the chiral symmetry breaking: the enhancement of the chiral condensate by the magnetic field will suppress the chirality density.

Figures.1-4 imply that the chirality density-density correlation has an important impact on the chiral magnetic effect, no matter the axial-vector interaction arises from the \I pairings, the OGEA, or the GCM.
The ratio $R$ can deviate significantly from the unity under the combined effects of $G_A$, $\mu_A$, $T$ and $\B$. The deviation may be positive or negative which depends on the sign of $G_A$. This contrasts with the role of $G_V$ on the CME: It only affects the CME obviously when the magnetic field $B$ is very large\cite{Fukushima:2010zza}; Its correction to the CME has nothing to do with $\mu_A$ and $T$.

Let us compare our results with the recent lattice QCD data. In \cite{Yamamoto:2011gk}, the CME was qualitatively confirmed by the lattice simulation via introducing a finite $\mu_A$. However, the obtained coefficient $C$ is around $0.013$, which is far less than the analytical value 1/${2\pi^2}$$\simeq$0.05. The further study using the quenched simulation suggests that the suppression is sensitive to the lattice spacing\cite{Yamamoto:2011ks}. The new lattice coefficient $C$ is roughly in the range 0.02-0.03 \cite{Yamamoto:2011ks}, which increases greatly compared to the former results. However, it still much less than the standard analytical value. Moreover, another new point in \cite{Yamamoto:2011ks} is that the coefficient $C$ might be dependent on the temperature.

Regardless the possible errors in \cite{Yamamoto:2011ks}, we stress that the current lattice data are quite natural when considering the effect of the attractive axial-vector interaction: (1) According to \eqref{eq:jz2f}, the induced electric current is proportional to the effective axial chemical potential $\tilde{\mu}_A$, which may be suppressed significantly compared to $\mu_A$ by the positive $G_A$. As demonstrated in Figs.1-4, the suppression is really remarkable for the moderate and strong couplings. So the reduction of $\tilde{\mu}_A$ by the attractive axial-vector interaction can explain why the lattice data of the induced electric current are much less than the analytical value. Let us make a rough comparison. According to Fig.7 in \cite{Yamamoto:2011ks}, the lattice ratio $C/2\pi^2$ is located in $0.4-0.6$ for $T\sim1.1T_c$($T_c\sim 270 \MeV$). Our numerical results shown in Fig.\ref{Fig:3} suggest that the ratio $R$ is in the range of $0.5-0.65$ for $T\sim1.1T_c$($T_c\sim 230 \MeV$) if $G_A/G_S$ varies from $1$ to $0.5$. We see that the lattice data are quite consistent with our results if the axial-vector interaction is attractive and strong enough. (2) The lattice data suggest that the coefficient $C$ or the ratio $C/2\pi^2$ is insensitive to $\mu_A$ and $e\B$. This feature is also in agreement with our results for the attractive axial-vector interaction, as shown in Fig.\ref{Fig:2} and Fig.\ref{Fig:4}. (3) One set of data ($\beta=5.90$ with $N_t=4,6,12$) in Fig.7 of \cite{Yamamoto:2011ks} indicates that the coefficient $C$ decreases with growing $T$. The author has argued that such a behavior might be a lattice artifact. However, that the induced current density $\jem$ may be suppressed by $T$ can be well interpreted by the attractive coupling $G_A$ as demonstrated in Fig.\ref{Fig:3}. According to Fig.7 of \cite{Yamamoto:2011ks}, the ratio $C/2\pi^2$ decreases from $\sim0.53$ to $\sim0.32$ when increasing $T$ from $317\MeV$ to $475\MeV$; Or in other words, the ratio $C/2\pi^2$ decreases roughly by $\sim0.04$ per $0.1T_c$. This is not so far away from the decline rate $\sim0.07$ per $0.1T_c$ for $G_A/G_S=0.5$, as indicated in Fig.\ref{Fig:3}.\footnote{ In the realistic case, the coefficient $C$ should change more mildly since the coupling $G_A$ becomes weaker with increasing $T$.} So to judge whether the coefficient $C$ is sensitive to the temperature or not needs more detailed lattice investigation.

On the contrary, the IIMM predicts that the CME current or the coefficient $C$ will be enhanced by the \I molecules. Clearly, this prediction deviates distinctly from the present lattice data. This might suggest that the dominant axial-vector interaction for $T>T_c$ is coming from the OGEA. Nevertheless, the current lattice simulation is still rough and far from conclusive(at leat at the quantitative level). First, that the current lattice data are sensitive to the lattice spacing \cite{Yamamoto:2011ks} suggests that the continuum limit is very important for quantitatively understanding the chiral magnetic effect. Second, the main simulations in \cite{Yamamoto:2011ks} are still at the quenched level while the discretization error of the fermion action is larger than that of the gauge action. Therefore, more improved quantitative lattice investigation will shed light on the role of the \I molecules on the chiral magnetic effect.

In any case, we stress that, due to the contribution of the dynamical axial chemical potential, the lattice result of the induced current for a finite $\mu_A$ doesn't need to be quantitatively consistent with the analytical formula. If no other QCD modifications, the coefficient $C$ or the ratio $R$ should be greater (less) than the standard analytical value for $G_A<0$ ($G_A>0$).

\subsection{Effect of axial-vector interaction on the $T-\mu_A$ phase diagram}

The chiral tricritical point (TCP) on the $T-\mu_A$ phase diagram has been found in \cite{arXiv:1003.0047}, which locates at the relatively larger $\mu_A$ region (in the chiral limit for vanishing quark number density). The TCP is also confirmed in the linear sigma model, whose location is very close to the $T$-axis\cite{arXiv:1102.0188}. These studies show that the $T-\mu_A$ phase diagram is somehow similar to the $T-\mu$ phase diagram of QCD.

\begin{figure}[t]
\begin{center}
\includegraphics[width=0.9\columnwidth]{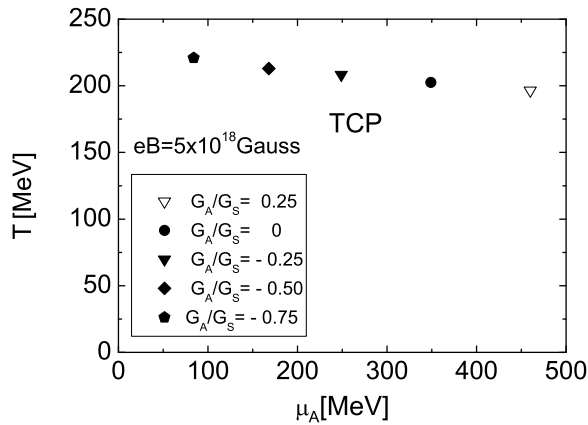} \vspace{1em}\\
\end{center}
\caption{\label{Fig:5} The locations of the chiral TCP in the $T-\mu_A$ plane for different $G_A$.}
\end{figure}

Here we will demonstrate the role of the axial-vector interaction on the $T-\mu_A$ phase diagram. We are particularly interested in its influence on the TCP. In Fig.\ref{Fig:5}, we show the locations of the TCP for different axial-vector interactions. We see that the TCP is quite sensitive to the axial-vector coupling. The negative $G_A$ shifts the TCP towards the smaller $\mu_A$ region while the positive $G_A$ moves it towards the larger $\mu_A$ area. This implies that the first-order chiral transition is strengthened by the repulsive axial-vector interaction induced by $I\bar{I}$ molecules while weakened by the attractive axial-vector interaction stemming from the OGEA.

It is well-known that the attractive vector interaction can soften the chiral transition and move the critical point(CP) towards the larger $\mu$ area via generating a dynamical quark chemical potential \cite{Asakawa:1989bq,Kitazawa:2002bc}. Especially, when taking into account the two-flavor color superconductivity, the vector interaction combined with the charge-neutrality constraint and/or the axial anomaly can even lead to multiple chiral critical points \cite{Zhang:2008wx}\footnote{The new critical point might also appear through the interplay between the chiral condensates and the diquark condensates in the color-flavor-locking phase in the context of the axial anomaly\cite{Hatsuda:2006ps}.}. We observe from Fig.\ref{Fig:5} that the attractive axial-vector interaction at finite $\mu_A$ has a similar influence on the chiral phase transition as the vector interaction does at finite $\mu$.

However, in the instanton molecule model, the axial-vector interaction at finite $\mu_A$ and the vector interaction at finite $\mu$ have opposite effects on the chiral transition. Since the induced $G_A$ and $G_V$ have different signs, the chiral transition will be strengthened (softened) at finite $\mu_A$ ($\mu$). In addition, the influence of the axial-vector interaction on the chiral transition at finite $\mu_A$ could be more remarkable since the  coupling $G_A$ is much stronger than $G_V$ in the instanton molecule picture.

Locating the CP on the $T-\mu$ phase diagram is a focus of the current studies on the QCD phase transition and heavy-ion collisions. So far, the location and even the existence of the CP is still under debate. Since there is no sign problem at finite $\mu_A$, it is very interesting to explore the $T-\mu_A$ phase diagram of the chirally imbalanced matter and locate the TCP by means of the lattice simulation. Such an investigation may provide useful information on the true QCD phase transition. Moreover, studying the $T-\mu_A$ phase diagram through lattice QCD may shed light on the non-perturbative features of sQGP. For example, we can test the instanton molecule picture of QCD by investigating the $T-\mu_A$ phase diagram.

\section{Summary and Conclusion}
\label{sec:Summary}
In this article, we have first investigated the influence of the axial-vector interaction on the chiral magnetic effect by introducing an axial chemical potential $\mu_A$. Such an interaction can be induced by the instanton molecules or derived from the one gluon exchange approximation of QCD. In the presence of a finite chirality density $n_A$, a dynamical axial chemical potential $\mu'_A$ is generated through the density-density correlation. We derived that the induced electric current density $\jem$ in an external magnetic field $\B$ linearly depends on the effective axial chemical potential $\tilde{\mu}_A$ which including the contribution of $\mu'_A$.
Accordingly, the obtained $\jem$ deviates from the standard analytical formula and explicitly depends on the chirality density $n_A$. In general, the deviation relies on the axial-vector coupling $G_A$, the axial chemical potential $\mu_A$, the temperature $T$, and the magnetic field $B$.

For $G_A>0$ from the one gluon exchange approximation, the induced current density can be significantly suppressed comparing with the analytical value. We also find that the ratio $R=\jem/\jem(G_A=0)$ is insensitive to the axial chemical potential and the magnetic field but decreases with increasing temperature if $G_A>0$. These features are quite consistent with the recent lattice results at finite $\mu_A$. Hence, the suppression of the lattice data comparing with the analytical formula can be attributed to the influence of the attractive axial-vector interaction of QCD. Actually, it is very likely that the deviation of the lattice result originates from the difference between $\mu_A$ and $\tilde{\mu}_A$. The reason is that Nielsen-Ninomiya's argument does not support the renormalization of CME if $\mu_A$ is not shifted dynamically.

On the contrary, an unconventional prediction from the instanton molecule model is that the axial-vector interaction is repulsive with $G_A<0$ and much stronger than the vector interaction. As a consequence, the chiral magnetic effect will be enhanced significantly by the instanton molecules near $T_c$. In such a picture, the ratio $\jem/\jem(G_A=0)$ increases with both $\mu_A$ and $T$. These features deviate obviously from the current lattice results. Since the present lattice simulation at finite $\mu_A$ is still rough and sensitive to the lattice spacing, we anticipate the more improved lattice calculation to test these predictions.

In addition, we also demonstrated that the axial-vector interaction plays an important role on the $T-\mu_A$ phase diagram of the chirally imbalanced matter. The repulsive (attractive) axial-vector interaction can effectively strengthen (weaken) the chiral phase transition if the coupling is relatively strong. The chiral TCP on the $T-\mu_A$ plane is shifted towards the lower (higher) $\mu_A$ region by the repulsive (attractive) axial-vector coupling. This is quite different from (similar to) the role of the vector interaction on the chiral phase transition at finite baryon chemical potential $\mu$.

Since there is no sign problem at finite $\mu_A$, the improved lattice simulation is still very welcome for disclosing the non-perturbative QCD effect on the chiral magnetic effect. Moreover, studying the deviation of the induced current density from the analytic result and locating the TCP in $T-\mu_A$ plane by the lattice simulation can be used to test the instanton molecule picture for $T>T_c$.

We note that if $\mu_A$ is introduced by coupling it only to the chiral density, it can't be considered as a true chemical potential. The reason is that the axial charge is not a conserved quantity due to the axial anomaly. Recently, it is proposed by Robakov \cite{Rubakov:2010qi} that $\mu_A$ should be conjugated to a proper combination of the chiral density and a Chern-Simon term, which is a conserved quantity. It is then claimed in \cite{Rubakov:2010qi} that the fact $\mu_A$ must be associated with a conserved charge is essential in the discussion for the nonrenormalization of the CME current.

Since there exists a subtlety on the definition of $\mu_A$, a question naturally arises that whether the dynamical axial chemical potential $\mu_A'$ defined in this paper can be associated with a conserved charge. We point out that $\mu_A'$ doesn't need to be conjugated  to a conserved charge if $\mu_A$ could be really well-defined. The reason is that $\mu_A'$ is an induced quantity generated dynamically by the axial-vector interaction for nonzero $n_A$. This is different from the parameter $\mu_A$ (Assuming it has been well-defined as a true chemical potential), which should be associated with a conserved charge in principle. We note that no matter how $\mu_A$ is introduced, it must give rise to nonzero $n_A$. As demonstrated in this paper, the $\mu_A'$ induced by $n_A$ will modify the chiral imbalance described originally by $\mu_A$.

A natural extension of this work is to investigate the role of the axial-vector interaction at finite $\mu$ in the presence of an external magnetic field. Conjugate to the induced current for the chiral magnetic effect at finite $\mu_A$, an  axial-vector current along the direction of the external magnetic field is generated at finite $\mu$\cite{Metlitski:2005pr}. Another related topic at finite $\mu$\ is the so called chiral shift parameter\cite{Gorbar:2009bm} in a magnetic field.  The roles of the axial-vector interaction in these cases will be reported elsewhere\cite{Zhang:future}.

\acknowledgments
The author thanks K. Fukushima and N. Yamamoto for useful discussions and helpful comments. This work was supported by the Fundamental Research Funds for the Central Universities of China.

\end{document}